\documentclass[conference,a4paper]{IEEEtran}

\usepackage[cmex10]{amsmath}
\usepackage{cite}
\usepackage{epsfig}
\usepackage{color}
\usepackage[usenames,dvipsnames]{xcolor}
\usepackage{nopageno} 
\usepackage{fancyhdr}
\usepackage[absolute]{textpos} 
\usepackage{float}
\usepackage{graphicx} 
\usepackage[T1]{fontenc}
\usepackage{textcomp}
\usepackage{caption}
\usepackage{subfigure}
\usepackage{array}

\usepackage{tikz}
\usepackage{lipsum}

\newcommand\copyrighttext{%
  \small 78\textperiodcentered 1-4799\textperiodcentered 6776\textperiodcentered 6115/\$31.00 \copyright 2015IEEE}
\newcommand\copyrightnotice{%
\begin{tikzpicture}[remember picture,overlay]
\node[anchor=south,yshift=10pt] at (current page.south) {\parbox{\dimexpr\textwidth-\fboxsep-\fboxrule\relax}{\copyrighttext}};
\end{tikzpicture}%
}

\begin{document}

\bstctlcite{IEEEexample:BSTcontrol}

\title{Enhancing Satellite System Throughput Using Adaptive HARQ for Delay Tolerant Services in Mobile Communications}

\author{\IEEEauthorblockN{Rami Ali Ahmad, J{\'e}r{\^o}me Lacan}
\IEEEauthorblockA{University of Toulouse, ISAE/DISC \& T{\'e}SA\\
Toulouse, France\\
Email: rami.aliahmad@isae.fr,\\ jerome.lacan@isae.fr}
\and
\IEEEauthorblockN{Fabrice Arnal, Mathieu Gineste}
\IEEEauthorblockA{Thales Alenia Space\\
Toulouse, France\\
Email: fabrice.arnal@thalesaleniaspace.com,\\ mathieu.gineste@thalesaleniaspace.com}
\and
\IEEEauthorblockN{Laurence Clarac}
\IEEEauthorblockA{CNES\\
Toulouse, France\\
Email: laurence.clarac@cnes.fr}}

\maketitle
\begin{abstract}

In this paper we propose the introduction of adaptive hybrid automatic repeat request (HARQ) in the context of mobile satellite communications. HARQ schemes which are commonly used in terrestrial links, can be adapted to improve the throughput for delay tolerant services. The proposed method uses the estimation of the mutual information between the received and the sent symbols, in order to estimate the number of bits necessary to decode the message at next transmission. We evaluate the performance of our method by simulating a land mobile satellite (LMS) channel. We compare our results with the static HARQ scheme, showing that our adaptive retransmission technique has better efficiency while keeping an acceptable delay for services.

\end{abstract}

\IEEEpeerreviewmaketitle


\section{Introduction}\label{sec:Intro}
\copyrightnotice
Link characteristics in mobile satellite communications make it difficult to exchange messages between transmitter and receiver(s). Long propagation delays (250 ms for geostationary satellite) is one of the main link characteristics in mobile satellite communications that can strongly affect the provided service. Another problem in mobile satellite communications is the transmission errors caused by the propagation impairments and the intra or inter-systems interferences.

Our objective is to propose a mechanism which optimizes the use of the bandwidth and thus improves the efficiency of link usage while providing an appropriate service to applications. The targeted services (data transfer from sensors, messages for aeronautical services, etc.) are assumed to be tolerant to delay. For example some aeronautical services define delay requirement for the delivery of 95\% of messages \cite{air_traffic}.

In this paper we propose to introduce adaptivity in HARQ mechanism. HARQ protocols are widely used in recent terrestrial wireless communication systems. Classical incremental redundancy (IR) HARQ, which transmits a fixed number of bits at each transmission, is not optimal from the efficiency point of view since sometimes the transmitter transmits more parity bits than needed to decode the message. An enhanced HARQ method to optimize classical IR HARQ in the context of mobile satellite communications was proposed in \cite{mypaper}. This method describes how to optimize the number of bits to be transmitted at each transmission, according to a predefined decoding probability table. However, adaptive retransmissions can improve the system performance and throughput level. Many papers have studied  adaptive retransmissions using adaptive coding modulation (ACM) and HARQ. Other contributions  proposed combinations of ACM or HARQ, where soft combining of transmission blocks in the receiver and block flat-fading channel are assumed \cite{Yu}. In \cite{Tarchi3} an approach based on the channel states for adaptive coding and modulation for mobile satellite communications is presented. Also ACM was studied with a multi-layer coding (MLC) in the forward link and open-loop adaptation in the return link in \cite{Arnau}. In \cite{Cioni} a deep analysis about channel estimation and physical layer adaptation techniques was made. Adaptive HARQ was also studied in \cite{Yongli} for a scheme based on punctured LDPC codes. An optimized IR HARQ schemes based on punctured LDPC codes over the BEC was proposed in \cite{Andriyanova}.

Contrarily to classical IR and enhanced HARQ \cite{mypaper}, the main idea of our adaptive HARQ technique is to send a feedback to the transmitter containing the estimated number of bits needed to decode the codeword with a targeted probability, if it has not been decoded successfully during the previous transmissions. This technique estimates the mutual information between the received and the sent symbols to predict the number of bits needed to decode at a predefined decoding probability for the next transmission. It uses the knowledge of the statistical distribution of the channel attenuation. The predefined decoding probability depends on the application/service (delay constraints). Some adaptive retransmission techniques based on mutual information have been studied \cite{Xiaoying}\cite{Xiang}, but not in a satellite communication environments neither using mutual information in combination with predefined decoding probability control. Our adaptive HARQ transmission proposal is simulated in a satellite communications environment, where an LMS channel and a  long round trip time are considered. 

The remainder of this paper is organised as follows. In Section II, we present LMS channel and its capacity. We present static HARQ in Section III. In Section IV, we propose an adaptive HARQ model for delay tolerant services in mobile communications. We present the results of simulations and we compare the performance of both techniques (static and adaptive HARQ) in Section V. 
We conclude our study in Section VI.

\section{Channel modelling}\label{sec:cahn_mod}
In our study on mobile satellite communications, we considered the LMS Channel to simulate this environment \cite{LMSPropa}\cite{LMSStatmodel}. In the following, we present the channel capacity and model. 

\subsection{Channel capacity and Mutual Information}\label{sec:chan_capa}

Channel capacity quantifies the maximum achievable transmission rate of a system communicating over a band-limited channel, while maintaining an arbitrarily low error probability.
It corresponds to the maximum of the mutual information between the input and the output of the channel, where the maximization is done with respect to the input distribution. Mutual information measures the information that $X$ (input of the channel, i.e. the sent symbols) and $Y$ (output of the channel, i.e. the received symbols) share. It is used in our approach to calculate the number of bits needed to decode a message at the next transmission with a given probability. For an equally distributed input probability, the mutual information, which corresponds to the capacity,  can be calculated by the equation given in \cite{mypaper}\cite{turbo}. 
%
%
%
%

For the rest of the paper, we define $MI_{req}$ as the average MI per bit required to decode a codeword for a given code at a given Word Error Rate (WER). 
The prediction of performance of the WER based on MI is quite classical, and has been described and validated in \cite{predic_perf}. 


Given the channel input symbol $x_i$ and its energy ${E_s}_i$ and a realisation of noise $n_i$ (has a Gaussian distribution with variance $\frac{N_0}{2}$), assuming perfect knowledge of the channel attenuation $\rho_i$, the channel output symbol $y_i$ can be written as:
\begin{equation}\label{input_output}
\begin{aligned}
y_i=\rho_i \sqrt{{E_s}_i}x_i+n_i\,.
\end{aligned}
\end{equation}


%

\subsection{LMS Channel}\label{sec:LMS_Chan}

There are many differences between the propagation on a terrestrial link and on a satellite link. One of the reference propagation models for the LMS channel is a statistical model based on a three state Markov chain \cite{LMSPropa}.
This model considers that the received signal originates from the sum of two components: the direct signal and the diffuse multipath. The direct signal is assumed to be log-normally distributed with mean $\alpha$(decibel relative to LOS(Line Of Sight)) and standard deviation $\Psi$ (dB), while the multipath component
follows a Rayleigh distribution characterized by its average power, $MP$(decibel relative to LOS). This model is called Loo distribution \cite{LMSStatmodel}\cite{Loo} . 
For the modelling of the LMS channel in our simulations, we use attenuation time series using a propagation simulator based on the three state channel \cite{LMSPropa}\cite{LMSStatmodel} provided by CNES. Using this tool we calculate the probability distribution function of the attenuation of the channel for a given environment. 








\section{Static Incremental Redundancy HARQ}\label{sec:first_scenario}
\begin{figure}[t]
\centering
\captionsetup{justification=centering}
\includegraphics[scale=0.33]{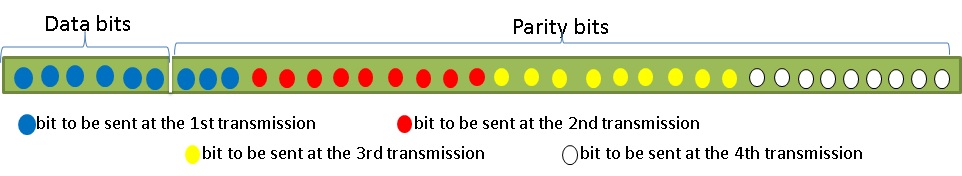} 
\caption{Example of the transmission technique considered in our simulations (Code rate 1/6, with a maximum of 4 retransmissions)}
\label{fig:puncturing}
\vspace{-10pt}
\end{figure}

In the classical incremental redundancy (IR) HARQ, the sender transmits a number of bits that correspond to a given codeword. After receiving the feedback (ACK/NACK) from the receiver, the transmitter decides to no longer send bits corresponding to this codeword if an ACK is received, or to send more parity bits if a NACK is received. 
The number of bits to be sent in the next retransmission is determined according to a table predefined at the sender. 
The parity bits are generated according to a coding scheme with a code rate corresponding to the maximum number of bits that can be transmitted per codeword. The bits to be sent at each transmission are part of the original codeword (mother code), leading to a different code rate at each transmission (see Fig. \ref{fig:puncturing}). This technique of transmission is somehow similar to puncturing. 

A configurable maximum number of transmissions is allowed for each message.

An enhanced HARQ method to optimize the number of bits to be transmitted at each transmission for IR HARQ was proposed in \cite{mypaper}. This method uses the mutual information concept and the knowledge of the global channel statistics to compute the number of bits leading to a given decoding probability. Decoding probabilities at each transmission are predefined and are chosen according to service and delay constraints.



\section{Adaptive HARQ for delay tolerant services in mobile satellite communications}\label{sec:second_scenario}
In this section, we propose an adaptive HARQ model using mutual information. Our method differs from the classical IR and enhanced HARQ \cite{mypaper}, by the feedback which contains an estimation of the number of bits still required by the receiver to decode the codeword if it is not decoded from the previous transmissions. The channel quality at the next transmission can not be predicted due to the long round trip time delay and the highly variable channel. The receiver computes the number of bits needed to decode the codeword with a probability of decoding predefined for each transmission. These probabilities are chosen according to delay constraints of the application.

Note that this model has some similarities with the models studied in \cite{Xiaoying} \cite{Xiang} which are also based on mutual information. However these models did not introduce predefined decoding probability at each transmission. In addition these models were not studied in mobile satellite communication environments. 

\subsection{Adaptive Retransmission Model} 
Our proposal is to compute the number of extra parity bits needed when the codeword cannot be decoded, using mutual information (MI). After receiving some bits of a given codeword, the receiver calculates the accumulated MI for this codeword. By considering a reference $E_s/N_0$, which corresponds to clear sky conditions (without attenuation), we suppose that the receiver can estimate the channel attenuation applied to these bits.
It assumes that the channel is stationary for the transmission time of the bits, at a given transmission for a given codeword. On each stationary time interval, the channel can be approximated by a gaussian channel. The MI obtained at the $j^{th}$ transmission for a given codeword can be computed as:

\begin{equation}\label{MI_1}
\begin{aligned}
MI^{(j)}=N_{sent}^{(j)}.MI((\rho^{(j)})^2.\frac{E_s}{N_0}),
\end{aligned}
\end{equation}

where:
\begin{itemize}
\item $\rho^{(j)}$ is the attenuation coefficient affecting bits transmitted at the $j^{th}$ transmission for a given codeword;
$(\rho^{(j)})^2$ is estimated by receiver measurement;
\item $N_{sent}^{(j)}$ is the number of bits transmitted at the $j^{th}$ transmission, affected by $\rho^{(j)}$;
\item $MI($\textperiodcentered$)$ is the function giving the value of mutual information for a given $E_s/N_0$ on a gaussian channel; 
\item ${N^{(j)}}$ is the total number of bits transmitted for a codeword up to the $j^{th}$ transmission.
\end{itemize}

The MI per bit accumulated for a given codeword, from the beginning of transmission until the $j^{th}$ transmission, can be computed as:
\begin{equation}\label{MI}
\begin{aligned}
MI^{(j)}_{acc}=\frac{N^{(j-1)}MI^{(j-1)}_{acc}+MI^{(j)}}{N^{(j)}}\,.
\end{aligned}
\end{equation}

By definition $MI^{(0)}_{acc}$=0.

Let us consider $MI_{needed}^{(j+1)}$ the minimum MI per bit needed to decode the codeword at the $(j+1)^{th}$ transmission with a predefined decoding probability. 

$MI_{req}$ is the average MI per bit required to decode a codeword for a given code at a given WER.

Then the number of bits $N_{needed}^{(j+1)}$ to be transmitted at the $(j+1)^{th}$ transmission can be obtained by the following equation:
\begin{equation}\label{needed_1}
\begin{aligned}
N_{bits}MI_{req}=N^{(j)}MI^{(j)}_{acc}+N_{needed}^{(j+1)}MI_{needed}^{(j+1)},
\end{aligned}
\end{equation}

Where $N_{bits}$ is the maximum number of bits that can be transmitted for a codeword. 

Finally $N_{needed}^{(j+1)}$ is given by:
\begin{equation}\label{needed_2}
\begin{aligned}
N_{needed}^{(j+1)}=\frac{N_{bits}MI_{req}-N^{(j)}MI^{(j)}_{acc}}{MI_{needed}^{(j+1)}}\,.
\end{aligned}
\end{equation}

$MI_{needed}^{(j+1)}$ is the key parameter for the computation of the number of bits to be sent at the $(j+1)^{th}$ transmission. In the following, we explain in detail how to proceed to calculate $MI_{needed}^{(j+1)}$ at each transmission according to the predefined decoding probabilities.

\subsection{Computation of $MI_{needed}^{(j+1)}$}
The LMS channel is a channel that changes quickly. Moreover, the round trip time in satellite communications is long. 
So, even if the receiver can estimate the channel quality for the last sequence of received bits, this information will not be useful at the transmitter considering the long delay and the highly variable channel. 

According to \eqref{needed_2} the value of $N_{needed}^{(j+1)}$ is computed from $MI^{(j)}_{acc}$ and $MI_{needed}^{(j+1)}$ which depends on the channel quality at the next transmission. Since this last value can not be known, we use the knowledge of the statistical distribution of channel attenuation to control the probability of decoding a codeword at each transmission. 

The idea is to define at the beginning of the communication a table containing the probability of decoding at each transmission. The decoding probability and the efficiency are related, the sender can transmit a large number of bits at the first transmission, which increases the decoding probability but the efficiency will decrease and vice versa. So, we have to improve efficiency while respecting delay constraints for services. 

Let $P_j$ be the probability of decoding at the $j^{th}$ transmission conditioned on the fact that decoding at earlier transmissions was impossible, where $P=\sum_j P_j$ is the percentage of decoded codewords over all the transmitted codewords. 


To target a decoding probability $P_j$ at the $j^{th}$ transmission, the receiver has to find the corresponding $MI_{needed}^{(j)}$.
$MI_{needed}^{(j)}$ corresponds to the mutual information received with an attenuation $\rho^{(j)}_{needed}$:

\begin{equation}\label{MI_req}
\begin{aligned}
MI_{needed}^{(j)}=MI((\rho^{(j)}_{needed})^2.\frac{E_s}{N_0})\,.
\end{aligned}
\end{equation}

Then to compute $MI_{needed}^{(j)}$, we have to find the attenuation coefficient $\rho_{needed}^{(j)}$ leading to the decoding probability at the $j^{th}$ transmission. 
The mutual information $MI($\textperiodcentered$)$ is a strictly increasing function (as a function of $\rho$). Then any attenuation coefficient greater than $\rho_{needed}^{(j)}$ will lead to a successful decoding. 
To determine $\rho_{needed}^{(j)}$ leading to $P_j$, we use the cumulative distribution function (CDF) of LMS Channel. 

To explain our calculation, we consider these two events:
\begin{itemize}
\item $A_j$: Successful decoding at the $j^{th}$ transmission;
\item $B_{j-1}$: Not decoding at the $(j-1)^{th}$ transmission.
\end{itemize}  
We observe that $P_j$ is equal to $p(A_j\cap B_{j-1})$. Let $p_j$ denotes $p(A_j)$. Since $A_j$ and $B_{j-1}$ are independent (according to the channel modelling),
\begin{equation}\label{proba_decod}
\begin{aligned}
P_j=p_j(1-\sum_{k=1}^{j-1}P_k),\\
\end{aligned}
\end{equation}

\begin{equation}\label{proba_decod1}
\begin{aligned}
p_j=\frac{P_j}{(1-\sum_{k=1}^{j-1}P_k)}\,.
\end{aligned}
\end{equation}

\begin{figure}[t]
\centering
\captionsetup{justification=centering}
\includegraphics[scale=0.4]{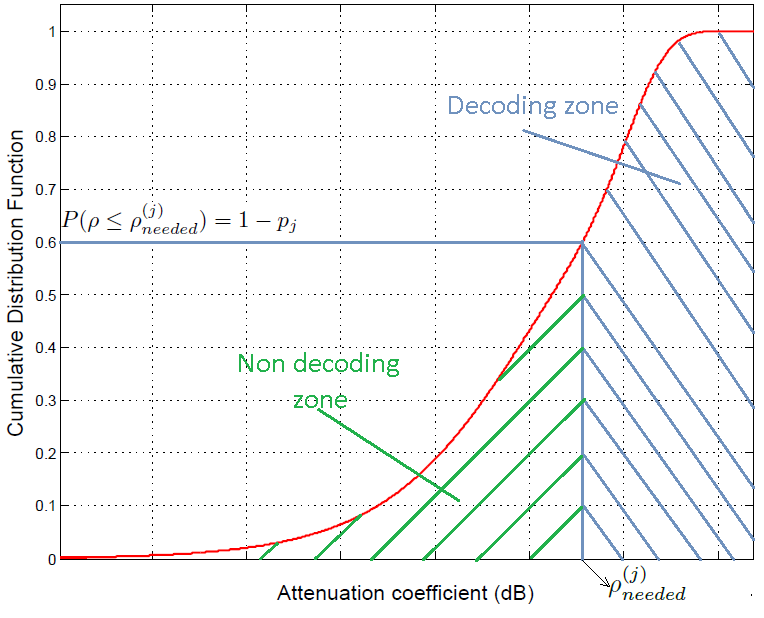}
\caption{The CDF of the attenuation coefficients in the LMS channel}
\label{fig:Cdf_its}
\vspace{-10pt} 
\end{figure}

The CDF of the channel gives us $P(\rho \leq \rho^{(j)}_{needed})$, while $p_j$ corresponds to $P(\rho \geq \rho^{(j)}_{needed})$ (successful decoding). Therefore, $\rho^{(j)}_{needed}$ on the CDF graph is given by 1-$p_j$ (See Fig. \ref{fig:Cdf_its}). 

So, let us summarize the different steps to find $MI_{needed}^{(j)}$. We first have $P_j$ from the table of predefined decoding probabilities. Then, we obtain $p_j$ from $P_j$ with \eqref{proba_decod1}. Then, we deduce $\rho_{needed}^{(j)}$ from $p_j$ with Fig. \ref{fig:Cdf_its}. Finally, we obtain $MI_{needed}^{(j)}$ from $\rho_{needed}^{(j)}$ with \eqref{MI_req}.

\begin{figure}[t]
\centering
\captionsetup{justification=centering}
\includegraphics[scale=0.35]{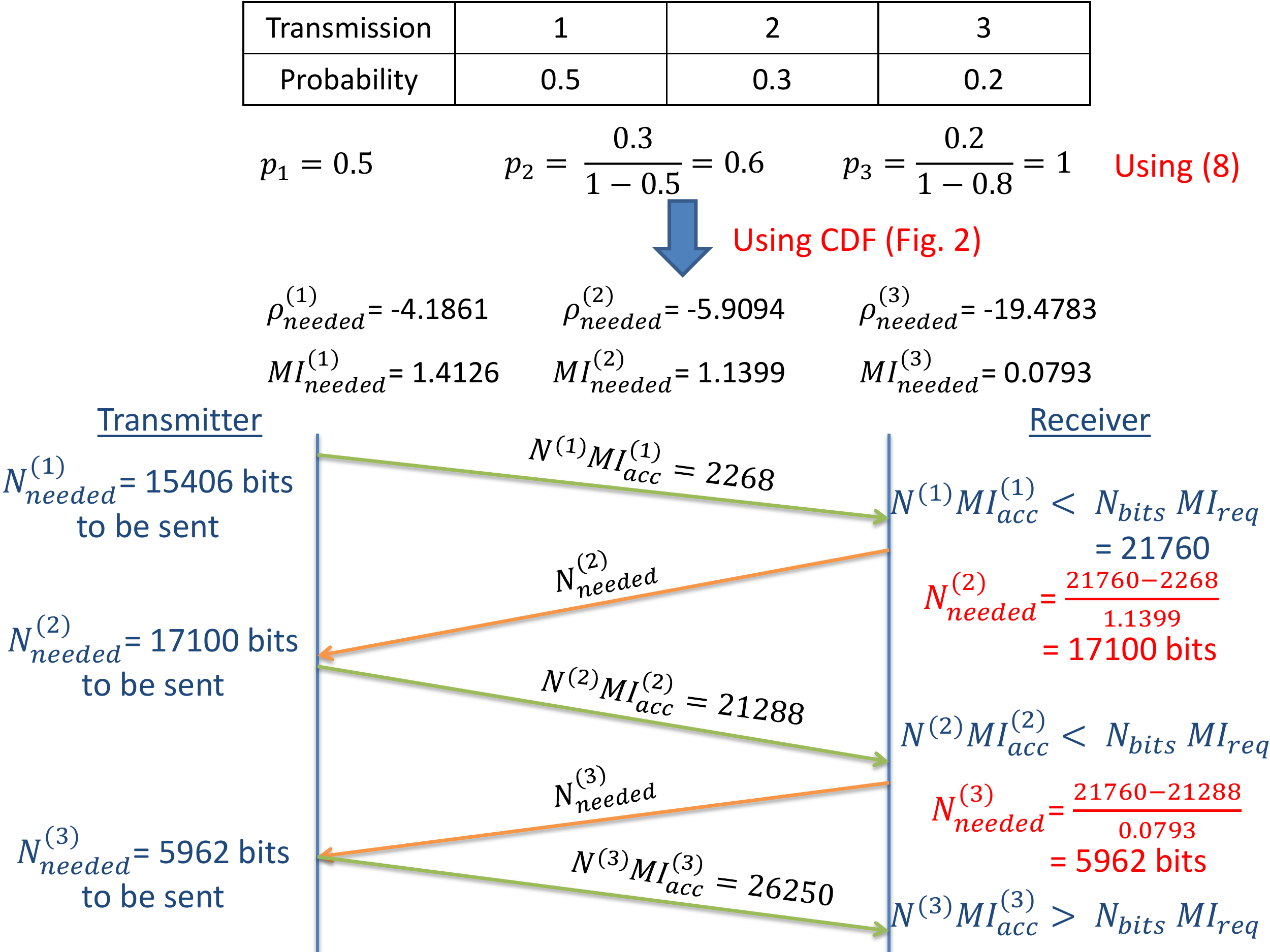}
\caption{Numerical example for the transmission of a given codeword within 3 transmissions ($E_s/N_0 = 7 dB$, code (8920,1/6), intermediate tree shadowed environment)}
\label{fig:example_trans} 
\vspace{-10pt}
\end{figure}

Fig. \ref{fig:example_trans} represents a numerical example for the transmission of a given codeword within 3 transmissions, showing different steps for the  calculation of all parameters in our model.



\section{Performance analysis} \label{sec:res}
System parameters considered for all simulations are:
\begin{itemize}
\item Satellite Orbit: Geostationary (GEO)
\item Round Trip Time: 500 ms
\item Band: S
\item Land Mobile Satellite Channel 
\item Speed of the mobile: 60 Km/h
\item Travelled distance: 10 Km
\item Mother FEC code, CCSDS Turbo codes 1/6
\item Codeword length: 53520
\item Modulation: QPSK
\item Symbol time: $4.10^{-6}$ seconds, bit rate (Rb): 500 Kbps 
\end{itemize}
Our simulations are about 10 minutes of communication between the transmitter and the receiver (about 300 Mb transmitted). In our simulations, we consider a targeted WER of $10^{-4}$, and we use the performances of CCSDS Turbo Codes (8920,$\frac{1}{6}$) as presented in \cite{Dolinar}. We suppose that the synchronisation is never lost. 
The transmitted codewords are identified with a sequence number that is never lost. We suppose also that the return channel does not introduce errors and the feedback can be transmitted immediately (no congestion problem on the reverse link).

The simulations use the computation of mutual information, as a real receiver would implement, according to the previously proposed algorithm. 
Doing this have required (in simulations) a calibration phase that has taken into account the actual numerical performances of the targeted FEC code(s). This allows to avoid implementing a real decoder in the simulation chain, while assuring a very good accuracy of the representation \cite{predic_perf}. To simplify, we decide whether the codeword is decoded or not using this formula:
 \begin{equation}\label{decod}
\begin{aligned}
N^{(j)}MI^{(j)}_{acc}\ge N_{bits}MI_{req}\,.
\end{aligned}
\end{equation}
 

\subsection{Impact of increasing the number of transmissions on the efficiency}
To show the importance of allowing many transmissions for the same codeword, we have considered three cases for our adaptive HARQ approach (explained in the previous section). In case 1, only one transmission is allowed, it corresponds to CCM (constant coding and modulation) technique. The transmitter transmits the same number of bits for all codewords, to decode from the first transmission. In case 2, two transmissions are allowed and in case 3, four transmissions are allowed. Their decoding probabilities are given in Table \ref{table:proba_threecases}.

Note that these probabilities do not represent values for specific applications. They are chosen arbitrarily to test our model.
The last value in the table is 0.9999 and not 1 because WER considered in our simulations is $10^{-4}$, then the total probability of decoding is $1-10^{-4}$.



\begin {table}[h!]
\begin{center}
\caption{PREDEFINED DECODING PROBABILITY TABLE FOR THE 3 CONSIDERED CASES}
\begin{tabular}{|c|c|c|c|c|}
  \hline
  Transmission & $1^{st}$ & $2^{nd}$ & $3^{rd}$  & $4^{th}$  \\
  \hline 
 $P_i (Case\, 1)$ & $0.9999$ & $-$ & $-$ & $-$  \\
 \hline 
 $P_i (Case\, 2)$ & $0.5$ & $0.4999$ & $-$ & $-$  \\
 \hline 
 $P_i (Case\, 3)$ & $0.5$ & $0.3$ & $0.1$ & $0.0999$  \\
  \hline
\end{tabular}
\label{table:proba_threecases}
\vspace{-1pt}
\end{center}
\end {table}

We define the efficiency, E (bits/symbol), as follows:
  \begin{equation}\label{spectral_efficiency}
\begin{aligned}
E=\frac{N_{data}\,\,_{bits} N_{decoded}\,\,_{words}}{N_{total}},
\end{aligned}
\end{equation}
where: 
\begin{itemize}
\item $N_{data}\,\,_{bits}$ is the number of data bits (useful bits) per codeword considered in our coding scheme;
\item $N_{decoded}\,\,_{words}$ is the total number of decoded codewords during the communication;
\item $N_{total}$ is the total number of symbols transmitted during the communication.
\end{itemize}

\begin{figure}[t]
\centering
\captionsetup{justification=centering}
\includegraphics[scale=0.6]{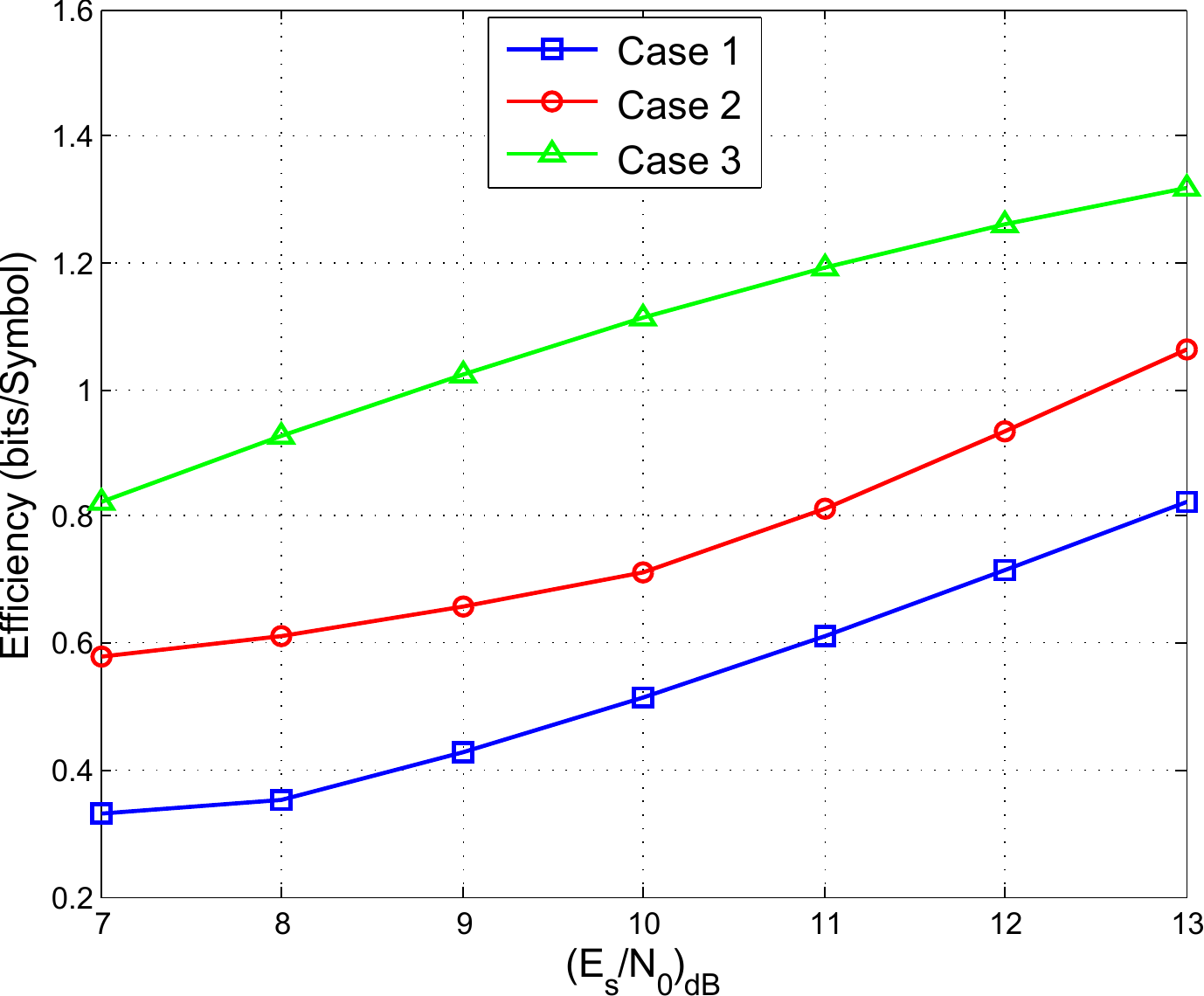} 
\caption{Efficiency obtained with three cases considered for different values of reference $(E_s/N_0)_{dB}$}
\label{fig:eff_cases}
\vspace{-10pt}
\end{figure}

Fig. \ref{fig:eff_cases} presents the efficiency obtained after simulating these three cases (with LMS channel, intermediate tree shadowed environment (ITS)) for different values of reference $E_s/N_0$ (NB: the channel variations $\rho^2(t)$ still being applied to this reference $E_s/N_0$, as explained in previous sections). This figure shows that case 3 (allowing four transmissions) outperforms case 1 and case 2 (allowing only one transmission and two transmissions respectively) in term of efficiency. This result can be interpreted by the fact that allowing only one transmission requires a large number of bits to be transmitted, sometimes more than the number of bits needed, to insure a complete decoding. When many transmissions are allowed, the number of bits to be transmitted at each transmission is optimized and computed in an adaptive way which optimizes the efficiency. 

\subsection{Simulations of static and adaptive HARQ}
We present results obtained by implementing three HARQ schemes (classical IR, enhanced HARQ proposed in \cite{mypaper} and our proposed adaptive HARQ) described in the previous sections, and we compare these results. Simulations were done with two types of LMS channel environments: ITS and open \cite{LMSPropa}.

We are interested to improve the efficiency while maintaining an acceptable delay for services. So, we will set the decoding probabilities to the same values for both models (enhanced HARQ and our proposed adaptive HARQ), while we set the number of bits to be transmitted at each transmission for the classical IR at the same value, and then compare the resulting efficiencies.

The number of bits to be transmitted at each transmission for the classical IR HARQ scheme can follow several strategies. As a simple case, we consider an equally shared repartition of the data+parity bits among the transmissions, as shown in Table \ref{table:static}.
\begin {table}[h!]
\begin{center}
\captionsetup{justification=centering}
\caption{NUMBER OF BITS TO BE TRANSMITTED AT EACH TRANSMISSION FOR CLASSICAL IR HARQ SCHEME}
\begin{tabular}{|c|c|c|c|c|}
  \hline
  Transmission & $1^{st}$ & $2^{nd}$ & $3^{rd}$  & $4^{th}$  \\
  \hline 
 $N^j_{sent} (bits)$ & $13380$ & $13380$ & $13380$ & $13380$  \\
  \hline
\end{tabular}
\label{table:static}
\vspace{-1pt}
\end{center}
\end {table}



Fixed decoding probabilities at each transmission considered in our simulations are given in Table \ref{table:proba_threecases} (case 3) where the maximum number of transmissions for a given codeword is four. It corresponds to a service accepting the delivery of 80\% of the messages at the first two transmissions and 20\% at the last two retransmissions. 

The delay for decoded codewords (at the receiver) can be expressed in terms of number of transmissions ($N_{trans}$), bit rate ($R_b$), number of bits sent ($N$) and propagation delay ($T_{propag}$), assuming a negligible access delay: 
\begin{equation}\label{delay}
\begin{aligned}
Delay= \frac{N}{R_b}+2(N_{trans}-1)T_{propag}+T_{propag}\,\,\, (s)\,.
\end{aligned}
\end{equation}

\begin{figure}[t]
\centering
\captionsetup{justification=centering}
\includegraphics[scale=0.6]{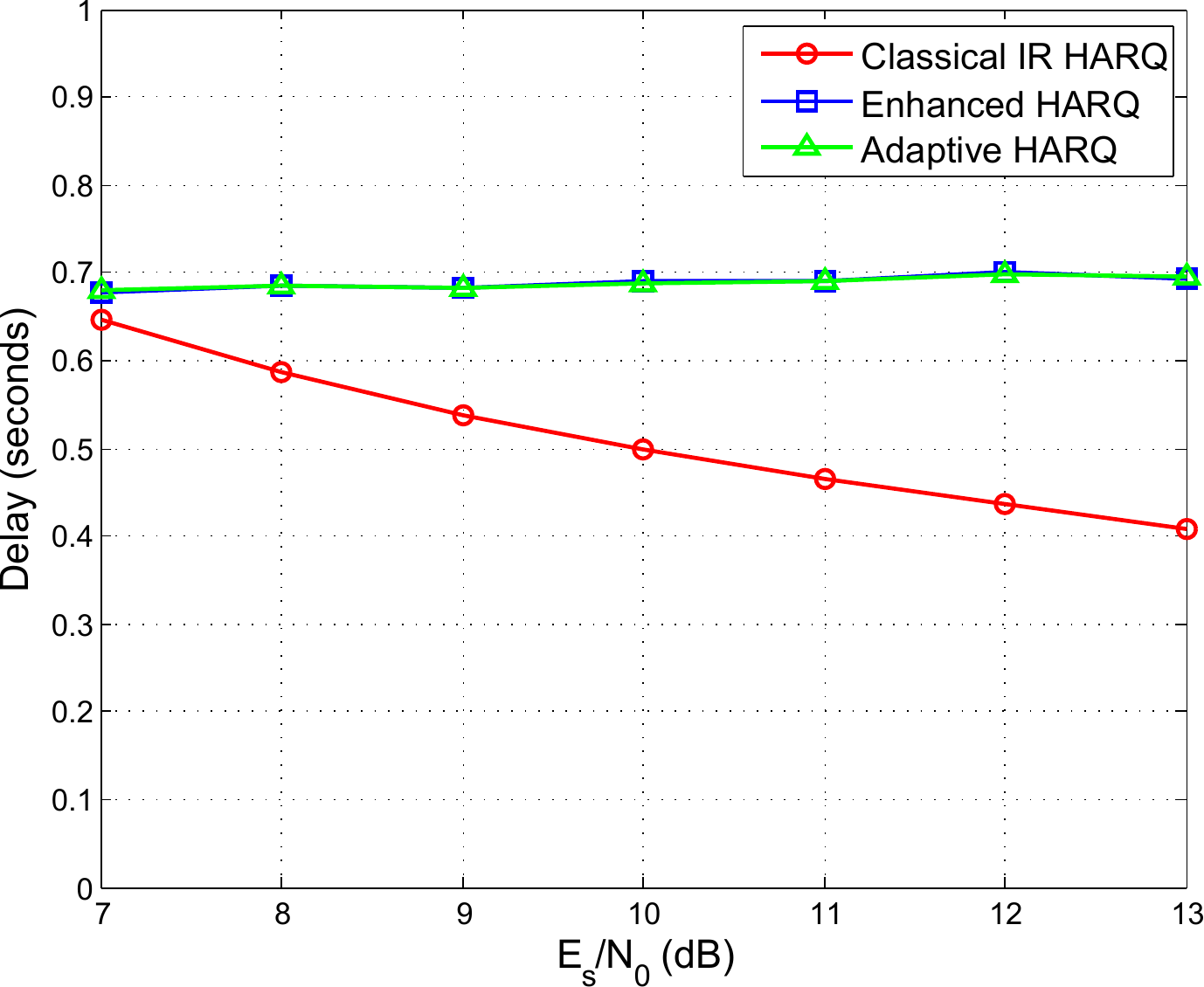} 
\caption{Delay for three schemes for different values of reference $(E_s/N_0)_{dB}$ (case 3) obtained in ITS environment}
\label{fig:delay_its}
\vspace{-10pt}
\end{figure}

Fig. \ref{fig:delay_its} shows the mean delay required to decode codewords, obtained for three schemes considered (by means of simulations with ITS environment). The mean delay is computed by averaging delays obtained for decoded codewords calculated using \eqref{delay}. As we can see, the values obtained for enhanced HARQ and adaptive HARQ are approximately the same. The delay is controlled by the decoding probability and remains globally constant. It changes a little, from a reference $E_s/N_0$ to another, according to the number of bits transmitted and the values of decoding probabilities obtained which are approximately the same as in Table \ref{table:proba_threecases} (case 3). However the delay obtained with classical IR HARQ decreases while the $E_s/N_0$ increases. Values of $E_s/N_0$ are chosen between 7 and 13 dB. According to the code and the environment considered, the percentage of decoding is not 100 \% for values of $E_s/N_0$ less than 7 dB.

\begin{figure}[t]
\centering
\captionsetup{justification=centering}
\includegraphics[scale=0.65]{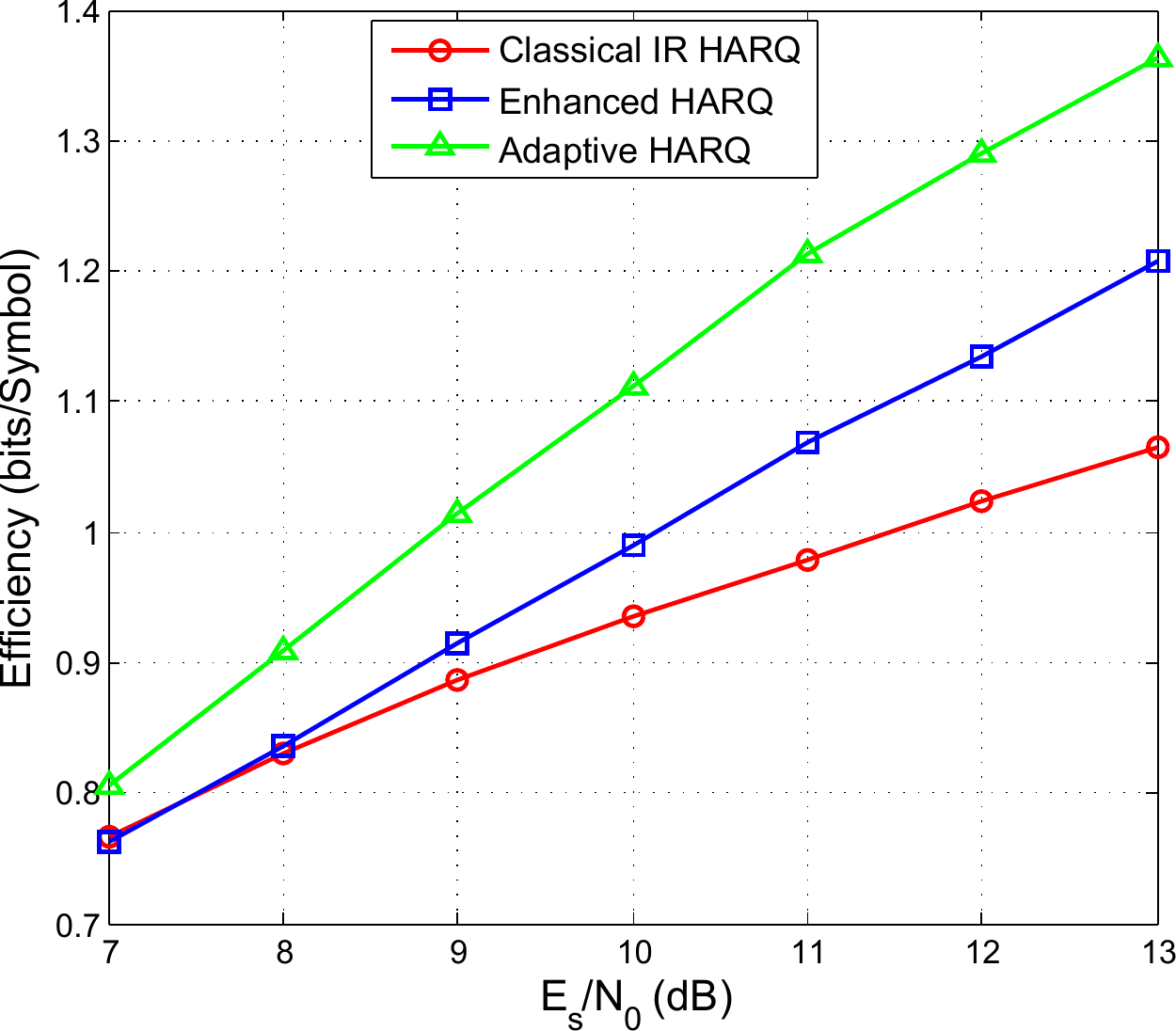} 
\caption{Efficiency for three schemes for different values of reference $(E_s/N_0)_{dB}$ (case 3) obtained in ITS environment}
\label{fig:eff_its}
\vspace{-10pt}
\end{figure}

To show the performance of the proposed adaptive model, Fig. \ref{fig:eff_its} compares the efficiency obtained with three schemes in the same conditions. We can see that the adaptive model outperforms the classical IR and enhanced HARQ especially for high values of reference $E_s/N_0$. Gain obtained is between 5\% and 13\% compared to the enhanced HARQ. However, it is between  5\% and 28\% compared to the classical IR. 
These results seem promising, however has been obtained with decoding probabilities chosen arbitrary. Therefore, some further improvements could probably be obtained by optimizing the decoding probability for each transmission step.

\begin{figure}[t]
\centering
\captionsetup{justification=centering}
\includegraphics[scale=0.57]{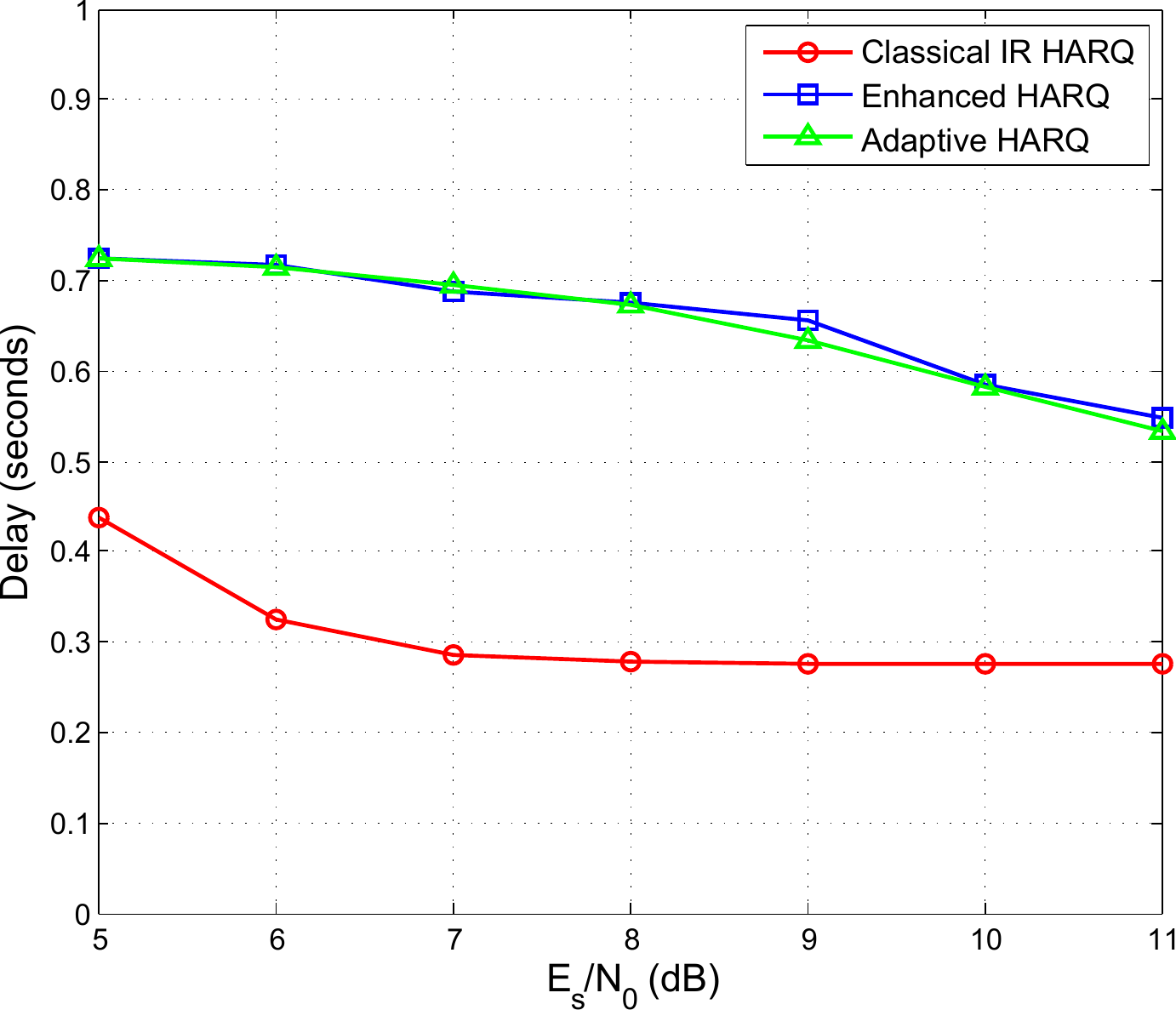} 
\caption{Delay for three schemes for different values of reference $(E_s/N_0)_{dB}$ (case 3) obtained in open environment}
\label{fig:delay_open}
\vspace{-10pt}
\end{figure}

In Fig. \ref{fig:delay_open} we present the delay obtained for the three schemes considered, simulated with open environment. In this type of environment, attenuations are not very high as in ITS environment. For this reason, at a given value of $E_s/N_0$, the number of bits estimated \eqref{needed_2} allows to decode at a probability more than that predefined because the mutual information reaches its maximum at 10 dB (QPSK modulation). 
For the case of classical IR HARQ, the delay reaches its minimum when the number of bits considered allows to decode all codewords from the first transmission.

\begin{figure}[t]
\centering
\captionsetup{justification=centering}
\includegraphics[scale=0.52]{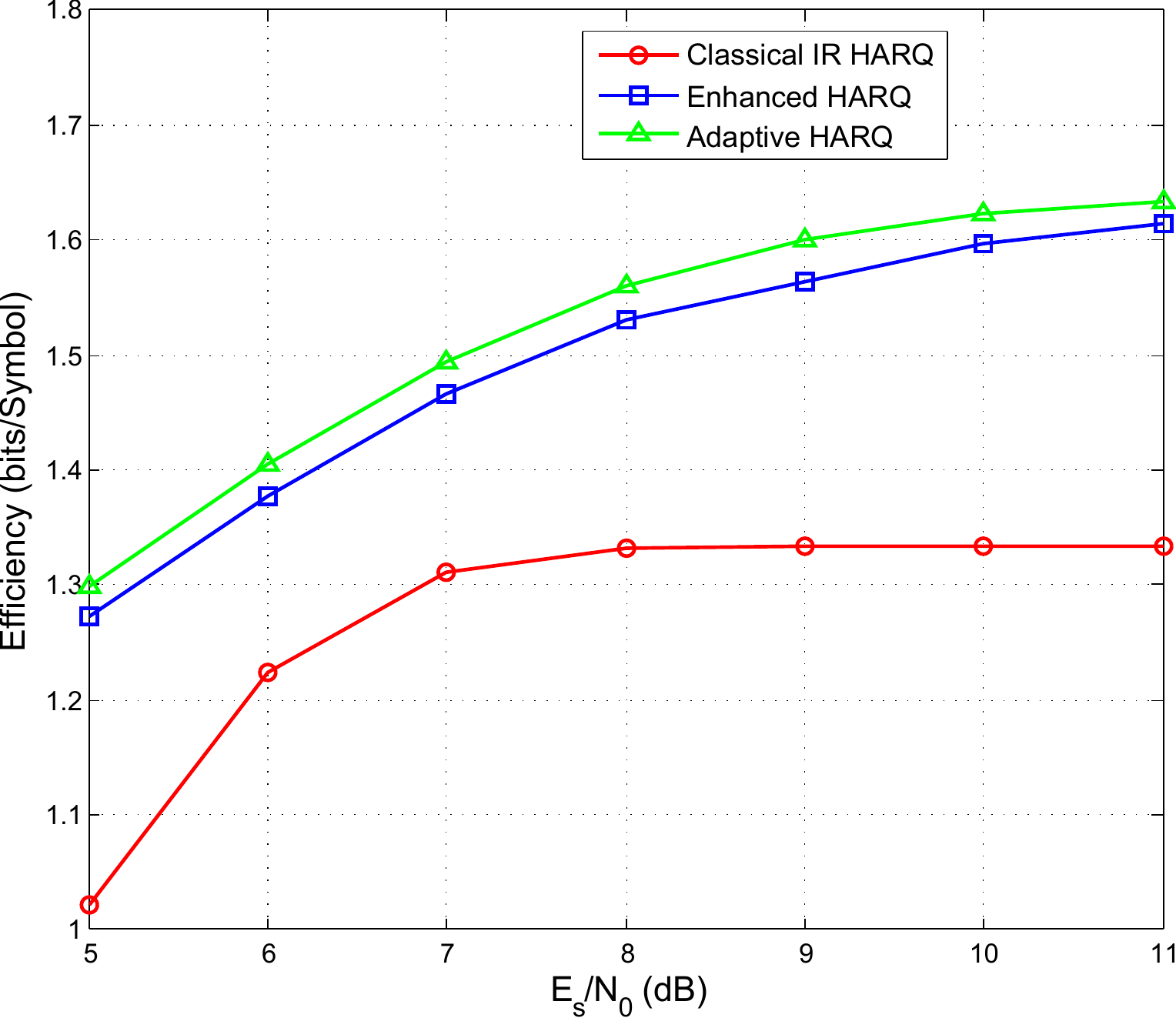} 
\caption{Efficiency for three schemes for different values of reference $(E_s/N_0)_{dB}$ (case 3) obtained in open environment}
\label{fig:efficiency_open}
\vspace{-10pt}
\end{figure}

Fig. \ref{fig:efficiency_open} presents efficiency obtained for three considered schemes, simulated with open environment. As we can see, classical IR HARQ reach its maximum when the number of bits considered is sufficient to decode from the first transmission. As in the case of ITS environment, adaptive model outperforms the classical IR and enhanced HARQ. However the gain obtained with open environment is about 2\% compared to enhanced HARQ and is between 14\% and 27\% compared to classical IR HARQ. These reduction of gain compared to that obtained with ITS environment is due to the fact that the attenuations in open environment are not very high as in ITS, as we explained above.

\section{Conclusion} \label{sec:con}

In this paper, we have compared three techniques of HARQ transmission for satellite communications. The first one is a classical IR HARQ scheme; the second one is an enhanced static IR HARQ, where the number of bits to be transmitted at each transmission is calculated according to a decoding probability; the third one is an adaptive technique which takes into account the channel quality at each transmission. This adaptive technique estimates at each transmission the number of additional bits needed to decode a codeword that could not have been decoded from the previous transmissions. This estimation relies on the mutual information given the decoding probability at each transmission and the knowledge of the channel distribution. Finally, results obtained after simulating three schemes in a mobile satellite communication environment are compared in terms of decoding probability (delay) and efficiency. Results show that the adaptive scheme has better performance in terms of efficiency especially for high values of reference signal to noise ratio, while maintaining an
acceptable delay for services.

As a future work, real system parameters will be considered at the physical layer (in particular framing and overhead) to estimate the performance of the mechanism in real systems. 

\bibliographystyle{ieeetran}
\bibliography{mybib}

\end{document}